# Charge gain via solid-state gating of an oxide Mott system

*Lishai Shoham\*, Itai Silber, Gal Tuvia, Maria Baskin, Soo-Yoon Hwang, Si-Young Choi, Myung-Geun Han, Yimei Zhu, Eilam Yalon, Marcelo J. Rozenberg, Yoram Dagan, Felix Trier, and Lior Kornblum\**

Lishai Shoham, Dr. Maria Baskin, Prof. Eilam Yalon, Prof. Lior Kornblum
Andrew and Erna Viterbi Department of Electrical and Computer Engineering, Technion – Israel Institute of Technology, Haifa 3200003, Israel
E-mail: lishai@campus.technion.ac.il   ;    liork@technion.ac.il

Itai Silber, Dr. Gal Tuvia, Prof. Yoram Dagan
Raymond and Beverly Sackler School of Physics, Tel Aviv University, Tel Aviv 6997801, Israel.

Soo-Yoon Hwang, Dr. Myung-Geun Han, Dr. Yimei Zhu
Condensed Matter Physics and Materials Science, Brookhaven National Laboratory, Upton, NY 11793, USA

Soo-Yoon Hwang, Prof. Si-Young Choi
Department of Materials Science and Engineering (MSE), Pohang University of Science and Technology (POSTECH), Pohang, Gyeongbuk, Korea 37673

Dr. Marcelo J. Rozenberg
Université Paris-Saclay, CNRS, Laboratorie de Physique des Solides, 91405 Orsay, France

Senior Researcher Felix Trier
Department of Energy Conversion and Storage, Technical University of Denmark (DTU), Kgs. Lyngby 2800, Denmark

**Abstract**

The modulation of channel conductance in field-effect transistors (FETs) via metal-oxide-semiconductor (MOS) structures has revolutionized information processing and storage. However, the limitations of silicon-based FETs in electrical switching have driven the search for new materials capable of overcoming these constraints. Electrostatic gating of competing electronic phases in a Mott material near its metal to insulator transition (MIT) offers prospects of substantial modulation of the free carriers and electrical resistivity through small changes in band filling. While electrostatic control of the MIT has been previously reported, the advancement of Mott materials towards novel Mott transistors requires the realization of their charge gain prospects in a solid-state device. In this study, we present gate-control of electron correlation using a solid-state device utilizing the oxide Mott system $La_{1-x}Sr_xVO_3$ as a correlated FET channel. We report on a gate resistance response that cannot be explained in a purely electrostatic framework, suggesting at least ×100 charge gain originating from the correlated behavior. These preliminary results pave the way towards the development of highly efficient, low-power electronic devices that could surpass the performance bottlenecks of conventional FETs by leveraging the electronic phase transitions of correlated electron systems.

## 1. Introduction

Field-effect transistors (FETs) have driven the computer and information revolutions by consistently enhancing performance and functionality while maintaining stable power consumption and cost per chip. However, silicon-based devices face inherent physical limitations, especially as they are scaled down to ultra-small dimensions. A key bottleneck is the reduction in switching efficiency due to short channel effects, which leads to performance degradation and increased power consumption.[1] To address these challenges, innovative materials are emerging as promising candidates for novel FET devices, with some offering electronic phase transitions that provide switching efficiencies unattainable in silicon.[2] Among these, oxide Mott systems hold exciting technological potential due to their ability to undergo metal-to-insulator transition (MIT) driven by strong electron correlations.[2,3] The capacity to control electron correlations via a gate electric field not only paves the way for innovative functional devices, but also provides a powerful tool for unveiling the underlying complex physics of strongly correlated electron systems, some of which are still under debate.

In Mott-Hubbard materials (early transition metals), the competition between the effective on-site Coulomb repulsion energy ($U_{eff}$) and the relatively narrow $3d$ bandwidth (W) gives rise to strong electron correlation ($U_{eff}/W$).[4,5] An electronic phase transition can be induced through electrostatic modification of the carrier density, thereby ultimately controlling the electrical resistivity via localization-delocalization of the channel electrons.[3] Such an MIT-based FET, often called a MottFET, offers the prospects for considerable resistivity modulation with small gate-induced perturbation.[6–8] Figure 1 presents a schematic example of gated MIT, where the Mott material constitutes the channel layer. At zero gate voltage, the channel remains in a metallic phase, allowing electrical conduction. In contrast, as the electrostatic charge induced in the channel by the gate approaches the integer-filling insulating state, the conduction electrons become heavier (more localized) due to the increase in electron correlation, thus decreasing the channel conductance.[9]

In analogy to MOSFET devices, we generalize the correlation between the electrons by defining an *effective* carrier density of the conduction electrons to be $n_{eff} = n/m^*$, where n and m* are the carrier density and effective mass of the electrons, respectively. It was predicted that small changes in carrier density induced by the gate (less than a factor of two between a correlated metal and a Mott insulator) in proximity to the MIT could lead to orders of magnitude changes in the effective carrier density, dramatically modulating channel conductivity.[2] This amplification of carrier density, termed "*charge gain*", is calculated to be up to ×100 greater than the charge induced by the gate.[10] This strong channel response to the gate is highly desirable for effective switching, constituting a key potential advantage of the MottFET.

Several approaches have been demonstrated for gating correlated oxides,[11–13] including electrolyte gating,[14–20] which is remarkable for fundamental studies, but faces practical limitations,[21] and photodoping, where light acts as a stimulus.[22–24] Focusing on Mott materials, solid-state gating of binary vanadates such as $VO_2$ has produced promising concepts,[25–28] yet challenges such as distinguishing electric field effect from thermal and filamentary effects, and temperature sensitivity remain.[29] Therefore, demonstrating a solid-state Mott-based field effect device with successful charge gain through gating is a critical milestone towards future electronics.

In this paper, we investigate the Mott-Hubbard type canonical system La$_{1-x}$Sr$_x$VO$_3$ (LSVO, $0.15 \leq x \leq 0.25$) around its filling control MIT.[30–33] The parent compound LaVO$_3$ (LVO) is a $3d^2$ Mott insulator with 2 electrons per unit cell (u.c.), and it undergoes an antiferromagnetic transition below $T_N$ = 140-160 K.[31] The other end compound, SrVO$_3$ (SVO), is a $3d^1$ correlated metal with 1 electron per u.c. and paramagnetic phase for all measured temperature range.[34] Substituting Sr$^{2+}$ (x) with La$^{3+}$ atoms affects the electron density n (chemical doping) by changing the V valance state, namely n = 2 - x.[31,35] Here, we demonstrate for the first time electric field control of electron correlation in perovskite vanadates. The results demonstrate a clear observation of effective mass enhancement through gate control, elucidating the high charge gain of the Mott-based channel by harnessing the strong electron correlation near the phase transition of the Mott-Hubbard system.

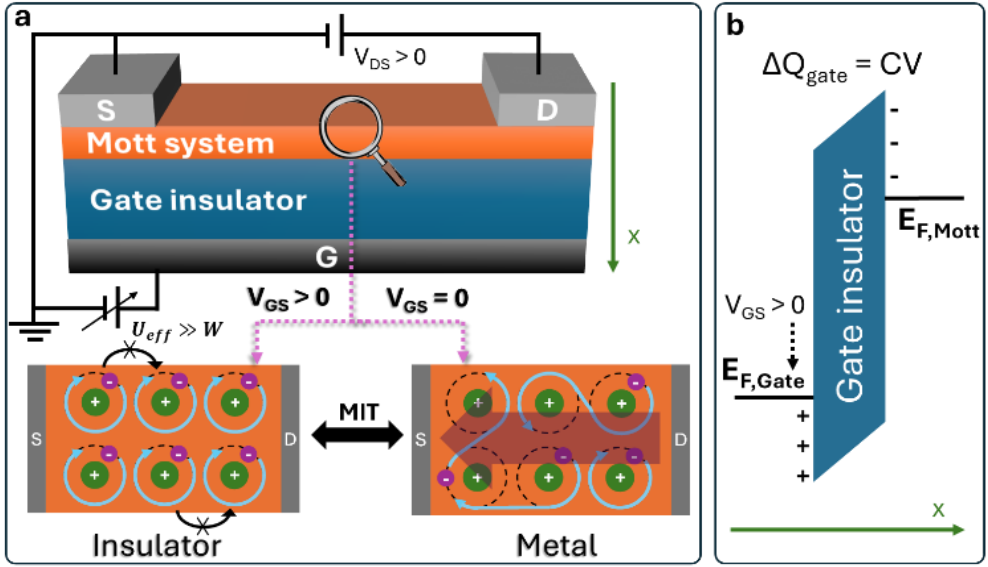

**Figure 1.** Schematic illustration of electrostatic modulation of the MIT. (a) A MottFET device. Under zero gate voltage ($V_{GS}$), the Mott-based channel remains in its metallic phase, allowing the conduction of electric current. Applying a positive gate bias dopes the channel with electrons, thus driving the Mott-based channel towards its integer-filling insulating state, where the electrons are localized, and the channel becomes electrically insulating. S, D and G represent the source, drain and gate, respectively. (b) A sketch of the electrostatics of the gate stack (capacitor) of the MottFET under positive gate voltage (the x-axis is defined in panel a). $\Delta Q_{gate}$ is the charge induced by the gate resulting from the gate insulator capacitance (C) and the applied gate voltage (V).

## 2. Results

To investigate the effect of electrostatic doping on an oxide Mott system, 10 u.c. LSVO films were grown on (001) SrTiO$_3$ (STO) substrates. The different Sr concentrations (x = 0.15, 0.20, and 0.25) were chosen around the filling-control MIT reported for LSVO single crystals at x = 0.18.[31] All films were capped *in situ* near room temperature with an ultrathin (~3 nm) oxide layer to protect their surface from over-oxidation (see Experimental for details).[36–38] At this Sr concentration range, the LSVO films are expected to be nearly lattice-matched to the substrate,[39,40] as evident by their out-of-plain lattice parameter (Figure S1a), indicating coherent growth and smooth surfaces (Figure S1b and S2, respectively). The LSVO/substrate interface exhibits an atomically abrupt interface (Figure 2a), which is particularly important for back-gated devices, as this is where the switching takes place.

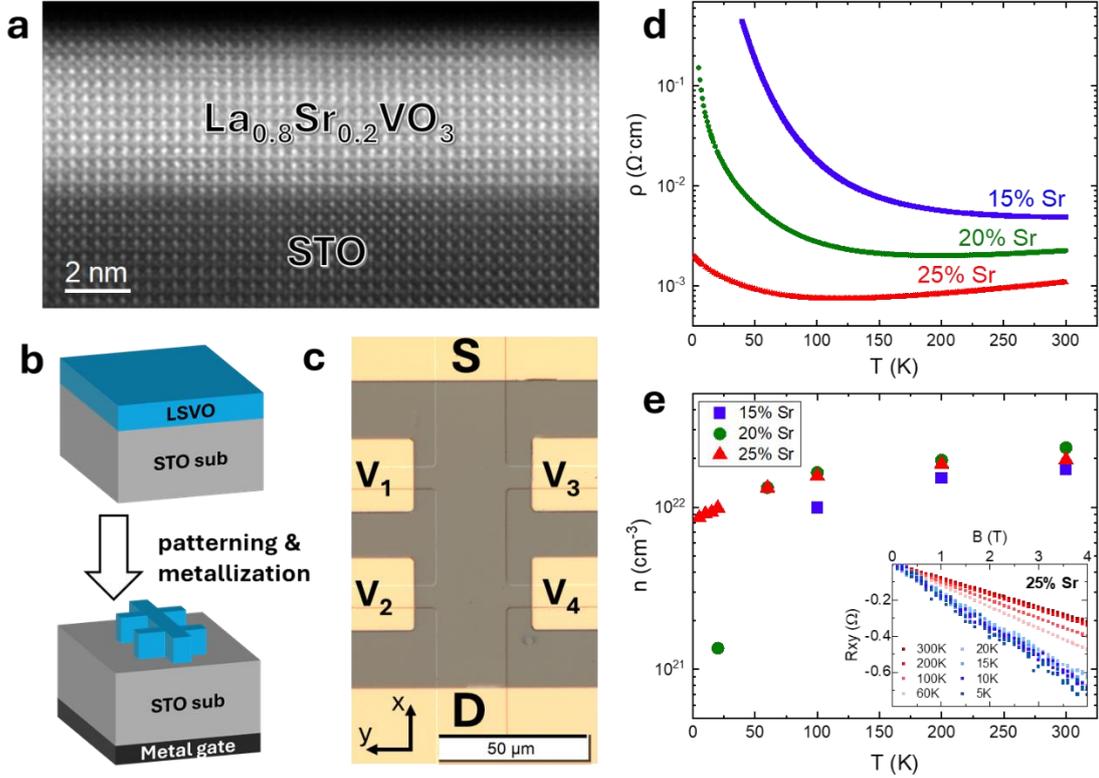

**Figure 2.** La$_{1-x}$Sr$_x$VO$_3$-based Hall bar devices. (a) Cross-section high-angle annular dark-field scanning transmission electron microscopy (HAADF-STEM) micrograph of a 10 u.c. La$_{0.8}$Sr$_{0.2}$VO$_3$ film on a 0.5 mm STO substrate (after device patterning). (b) A sketch of the Hall bar device, which is measured using a back gate configuration where the STO substrate constitutes the gate insulator. For clarity, an ultrathin oxide capping layer on top of the LSVO film is not shown. (c) Optical image of a Hall bar device in top view. The electrons flow from the source (S) to the drain (D), and the voltage is measured between V$_1$ and V$_3$ (or V$_2$ and V$_4$) for Hall voltage, and between V$_1$ and V$_2$ (or V$_3$ and V$_4$) for sheet resistance. (d) Temperature-dependent resistivity of the different LSVO-based devices, calculated with the nominal film thickness of 4 nm (10 u.c.). (e) Variation of the electron density at different temperatures as extracted from Hall measurements. We assume $n \approx n_H \approx (R_H e)^{-1}$ for simplicity (see supplementary materials for discussion). The inset presents the Hall response (R$_{xy}$) as a function of the magnetic field for the 25% Sr device at different temperatures.

2.1 Zero-bias behavior

To understand the effect of gating on the Mott channel, we start with a baseline analysis of the LSVO properties under zero gate bias. We patterned our films into Hall bar geometry,[41] as schematically and optically illustrated in Figure 2b and c, respectively. Figure 2d presents the temperature-dependent resistivity of the 10 u.c. thick LSVO films. The 15% Sr concentration device shows an insulating behavior (dρ/dT < 0), while for the 20 and 25% Sr concentration devices, the resistivity shows an upturn at ~200 K and ~120 K, respectively. Although the films are ultrathin, their resistivity behavior closely resembles that reported by Miyasaka et al. for LSVO single crystals.[31] Moreover, a comparison to the resistivity behavior of a "thick" (27 nm) 25% Sr concentration LSVO film shows similar values, thus indicating negligible surface scattering in the ultrathin films (Figure S3).

Hall measurements were performed to estimate the carrier density at various temperatures (Figure 2e). For the sake of simplicity, we consider the carrier density as $n \approx n_H \approx (R_H e)^{-1}$, where $e$ is the electron charge, and $R_H$ is the Hall coefficient. The negative value of $R_H$ for all measured devices indicates that the carriers are electrons. At room temperature, the extracted carrier density is of the order of $10^{22}$ cm$^{-3}$, in agreement with the expected 1-2 electrons per u.c. for all LSVO devices. The decrease in electron density for the 20 and 15% Sr concentration devices was previously observed in LSVO crystals, and it is attributed to the shrinkage of the Fermi surface as LSVO approaches the paramagnetic to antiferromagnetic magnetic phase transition.[31]

At the 100-300 K temperature range, the Hall response is linear with respect to the field ($R_{xy}$-B) for all samples, and no magnetic or structural phase transitions are expected.[31] When the Sr concentration in the LSVO film is decreased, it corresponds to the increase of the electron density (electron doping). Since LVO is an integer-filling Mott insulator, the decrease of Sr concentration drives the LSVO towards its Mott insulating state, as evident from the systematic resistivity increase visible in Figure 2d. For a given temperature, the comparison between the resistivity and carrier density upon electron doping (x) suggests that the origin of the resistivity change in response to electron doping is the result of mobility variation rather than carrier density (Figure S4).

2.2 Gating the Mott channel

Having established the zero-bias behavior, we turn to the electric field effect measurements on the LSVO devices in a back gate configuration. Figure 3a presents the percentage of the resistance change with respect to the zero-bias resistance, $R_0$ (Figure 3b), as a function of the back gate voltage measured at 100 K. The temperature was chosen to utilize the high dielectric constant of the STO substrate[42] without approaching the LSVO structural and magnetic phase transitions. We note the absence of hysteresis in the bi-directional sweeps, indicating no significant contribution of ionic or filamentary mechanisms. Taking the thickness of the gate insulator ($t_{STO}$= 0.5 mm) and its dielectric constant ($\varepsilon_{STO}(100K)$ = 1280, which does not change significantly with electric field in the range employed here[42]) we estimate the sheet charge modulation in the channel, as induced by the gate, $\Delta Q_{gate}$, via the equation:

(1)   $\Delta Q_{gate}(V_g) = V_g \cdot \varepsilon_0 \varepsilon_{STO}/t_{STO}$

Where $V_g$ is the applied gate voltage, and $\varepsilon_0$ is the vacuum permittivity. As shown in the top x-axis in Figure 3a, applying 200 V on the gate yields a maximum charge variation of $\sim 3\times 10^{12}$ cm$^2$; this should be considered an upper bound, as it neglects possible defects and traps in the substrate that could screen the gate voltage.

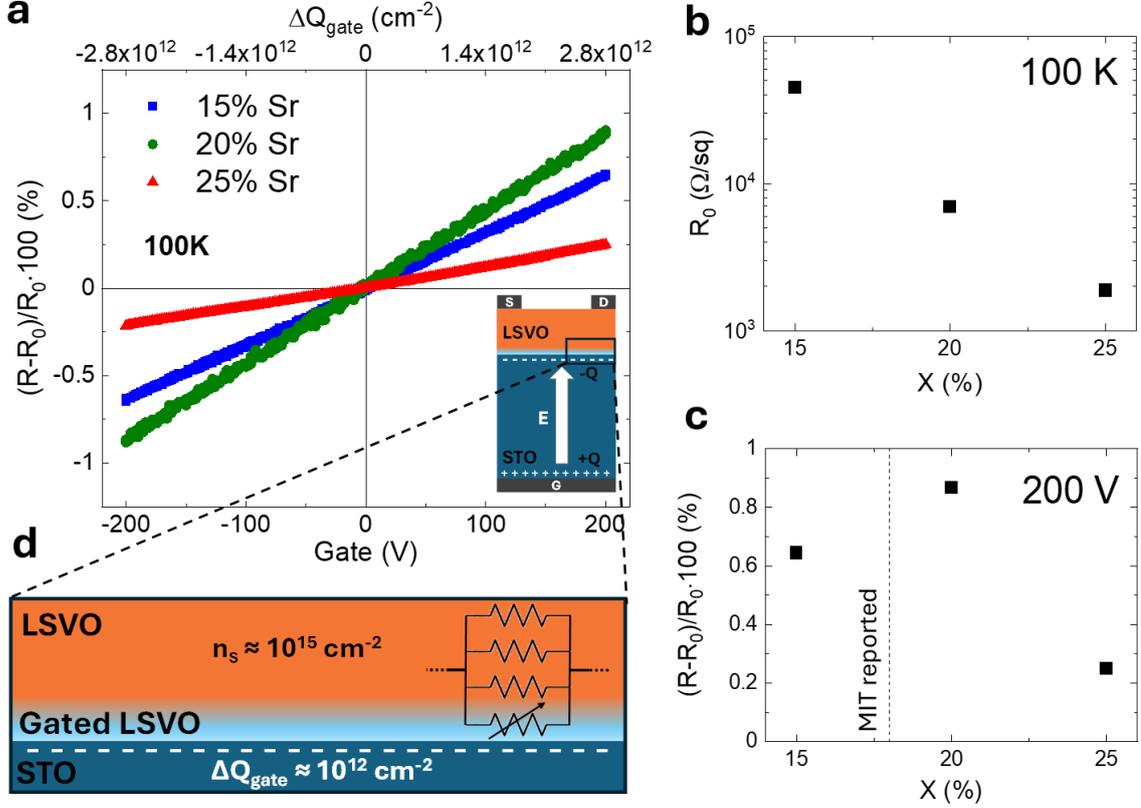

**Figure 3.** Gating a Mott channel. (a) Resistance variation as a function of the gate voltage for the different LSVO devices measured at 100 K, where the resistance R is measured in a 4-point configuration (Hall bar), and $R_0$ is the resistance measured at zero bias. The scan of the gate voltage is a bi-directional sweep. The upper x-axis represents the calculated electron charge induced by the gate (Eq. 1), where $\Delta Q_{gate} > 0$ indicates electron density increase. The inset shows a schematic illustration of the back gate measurement in a cross-section where positive voltage is applied on the gate (G), thus increasing the electron density in the LSVO channel. The measured current is between the source (S) and the drain (D), where an optical image of the device in a top view is presented in Figure 2c. (b) and (c) present the zero bias resistance (b) and resistance variation under 200 V gate voltage (c) as a function of Sr concentration (x) at 100 K. (d) A "zoom-in" sketch of the gated channel illustrating the change in resistance where the LSVO channel was affected by the gate voltage (the total measured film resistance is the sum of parallel resistors). The ×1000 difference in sheet carrier density between the $\Delta Q_{gate}$ and the LSVO conduction electrons, whereas the resistance changes by ×100, pinpoints the charge gain potential of the Mott-based channel.

Remarkably, all LSVO devices exhibit an unusual resistance response to the gate voltage. Considering that this is an electron system, the channel resistance is expected to decrease with the increase of electron density; however, we observe a resistance *increase* with the increase of electron density. This behavior suggests that the increase in electrons within the channel triggers an additional effect that surpasses the impact of merely increasing carrier density. This is the key result of this work, showcasing the expected behavior of a MottFET device as discussed below.

**3. Discussion**

3.1 Effective mass enhancement

We attribute the resistance increase with increasing gate voltage to the behavior of a Mott system approaching its insulating state, similar to what is "statically" achieved through chemical doping. In general, MIT can be described as $n/m^* \to 0$, suggesting that although the electron concentration (n) increases, the increase in the effective mass ($m^*$) outweighs it, as was theoretically predicted for a Mott system approaching its integer-filling insulating state.[9] In other words, $m^*$ increases enough to counteract the increase in n, causing $n/m^*$ to decrease and resistance to rise. The careful analysis of our experimental results strongly points to the long sought-after evidence of a divergence in the effective mass in the Mott channel driven by gate control. This observation may not only validate theoretical predictions[9] but also achieves this in a solid-state device, an important milestone towards the technological control of the correlated phenomenon.

Figure 3c exhibits an apparent optimum in the resistance response to the gate voltage for the 20% Sr concentration device, which is likely due to its proximity to the MIT observed by Miyasaka et al. (18%).[31] This observation suggests the possibility of fine-tuning the effect to optimize the response and consequently device performances. We point out that the resistance response is not necessarily expected to be linear, and we attribute the linear behavior observed in Figure 3a to the small resistance range probed here.

3.2 Charge gain

These findings not only enhance our understanding of electron correlation effects in Mott materials, but also set the stage for future innovations in device engineering. To illustrate its potential for electronic devices, we make an analogy of the correlation effect to the operation of a MOSFET device by considering the gate voltage effect on the *effective* carrier density ($n/m^*$). Thus, although theoretically applying positive gate voltage is expected to increase the conduction electron density, the mass enhancement localizes the carriers. Therefore, there are significantly fewer delocalized electrons available for conduction, and the system is approaching its Mott insulating state. The comparison between the charge in the channel induced by the gate ($\Delta Q_{gate}$) and the conduction electrons "capable" of conduction is termed the charge gain of the system.

With the aim of quantifying the charge gain in the gate-induced resistance response, we first treat the channel in a 2-dimensional (2D) limit, assuming uniform electron density in the LSVO film. This scenario represents the (unlikely) case that the gate field affects the entire LSVO thickness, constituting a *lower bound* to the charge modulation. In this limit case, we can calculate the LSVO sheet carrier density ($n_s$) by multiplying n with the nominal film thickness of 4 nm, yielding a sheet carrier density on the order of $10^{15}$ cm$^{-2}$ for all LSVO devices at 100 K. We recall that the sheet charge variation induced by the gate ($\Delta Q_{gate}$) is calculated to have an upper bound of ~$3\times10^{12}$ cm$^{-2}$ at 100 K (Sect. 2.2). Thus, the gate is able to modulate merely 0.1% of the huge LSVO sheet carrier density of $10^{15}$ cm$^{-2}$; however, we are measuring up to 1% resistance increase under this voltage. Therefore, within a naïve rigid band (non-correlated) picture, such an observation would require a ×10 charge gain. The absence of a non-correlated framework to account for this result pinpoints the unique underlying physics of the Mott system and its promising potential for electronic devices.

The 2D limit discussed in the previous paragraph is an over-simplification, and the actual charge gain in this system is much higher. Since this is a 3D system, the measured resistance is an average quantity over the entire thickness of the LSVO film (sketch in the inset of Figure 3d). The shallow extent of the charge distribution (screening length) induced in the channel by the gate electric field is estimated to be ~ 1-2 u.c. (or less) out of the 10 u.c. film,[43,44] arising from the high LSVO carrier density. As schematically illustrated in Figure 3d, we can coarsely divide the LSVO film into (i) a shallow region that is affected by the gate voltage (termed "Gated LSVO"), where its resistance is modulated by the gate; and (ii) the remainder of the film, unaffected by the gate voltage due to the screening of the electric field by the free carriers. In other words, the remainder of the film is a parallel resistor to the gated LSVO layer, significantly obscuring the charge modulation in the latter. Taking this into account suggests at least another order of magnitude to the charge gain, to be in the ×100 regime. This demonstration of charge gain in a Mott-based field effect device highlights the potential of Mott materials for novel oxide electronics based on electrically controllable phase transitions.

## 4. Conclusion

In this work, we demonstrate the potential of a Mott-Hubbard system for novel functional devices by controlling its electron correlation with a solid-state gate. The resistance response of the Mott-based channel under gate bias suggests the long sought-after evidence of effective mass divergence near the integer-filling Mott insulating state. Moreover, the analogy to a MOSFET device highlights a few orders of magnitude charge gain, which is highly desired for enhancing device performance. Additional functionality and performance gain can be achieved through the integration of perovskite oxides with common semiconductors, either by direct epitaxy[45] or via epitaxial liftoff,[46,47] and with ferroelectric gates.[48–51] The gate measurements in this study were conducted at 100 K merely to utilize the high dielectric constant of the STO substrate; however, we expect similar behavior of the Mott channel at higher temperatures, thus allowing device operation at room temperature with a top gate device configuration. Channel thickness reduction and optimization are required to harness the full potential of the device, where the channel thickness is in the order of the gate screening length but still maintains negligible surface scattering. These findings not only demonstrate the feasibility of using Mott materials in advanced electronic applications but also provide a direct probe for understanding the gate control of electron correlation. The results further plot an initial design space for crafting these physics into novel functional devices.

## 5. Experimental Section

*Sample preparation:* Ultrathin (10 u.c.) LSVO films were epitaxially grown on (001)-oriented STO substrates (5×5×0.5 mm, CrysTec GmbH). The growth was done using oxide molecular beam epitaxy (MBE, Veeco GenXplor) at a background molecular oxygen pressure of ~5×10$^{-7}$ Torr and substrate (thermocouple) temperature of 1000 °C.[52] The Sr, La, and V atomic fluxes were independently calibrated in a vacuum prior to the growth using a quartz crystal microbalance (QCM) to a 1 u.c./min growth rate. Growth was done in a shuttered mode.[36,53] After the deposition of the LSVO film, the samples were cooled to room temperature under the same oxygen pressure. Once the samples reached temperatures <100 °C, protective ultrathin (~3 nm) oxide capping layers of $TiO_{2-x}$ or $Al_2O_{3-x}$ were deposited at an oxygen backpressure of 8×10$^{-7}$ Torr (a high temperature source and an e-beam source were used for the evaporation of Ti and Al, respectively). The LSVO growth started with 1 monolayer of SrO (following Hotta et al.[54]) to circumvent the possibility of 2DEG formation at the STO surface; further validation to the absence of 2DEG is via magnetotransport (Figure S5 and discussion therein).

*Device fabrication:* The LSVO films were patterned into Hall bar devices with the dimensions of 130×30 μm for the main bar, and 10×10 μm for the Hall bar legs. The contacts to the Hall bar are made from 40 nm Ti and 200 nm Au deposited using e-beam evaporation. Details about the Hall bar fabrication and metal contact deposition can be found elsewhere.[41] For the back gate electrode, a 100 nm aluminum contact was deposited using e-beam evaporation, then the sample was glued with silver paint to the measurement puck. The resistance between different Hall bar devices on the same sample was unmeasurably high, indicating negligible substrate conductivity.

*Scanning Transmission Electron Microscopy:* The cross-sectional TEM sample was prepared by focused ion beam (FIB) with 5 keV Ga ion for final milling. The STEM HAADF image was acquired using JEOL ARM 200CF with double spherical-aberration correctors operated at 200 kV. The collection angle was in the range from 68 mrad to 280 mrad.

*Device characterization:* Hall resistance and longitudinal resistance measurements were done using a Quantum Design Physical Properties Measurement System (PPMS). The carrier density was extracted from the slope of transverse resistance versus magnetic field sweep at each temperature (see supplementary materials for the data). We note that at approximately 50 K and 15 K for 15 and 20% Sr concentration devices, respectively, the Hall response ($R_{xy}$) became non-linear with the magnetic field (B). This non-linearity could not be reasonably explained by standard two-band fit, and may be a result of an antiferromagnetic anomalous Hall effect;[55] for simplicity, these data points are excluded from Figure 2e (see supplementary materials for details). The optical image of the Hall bar was taken using a Nomarski microscope with cross polarization.

**Supplementary materials**

See supplementary materials for additional film characterization, a detailed discussion about the magnetotransport analysis and gate measurements at low temperatures.

## Acknowledgments

This work was funded by the Israeli Science Foundation (ISF Grant No. 1351/21). F.T. acknowledges the support by research Grant No. 37338 (SANSIT) from Villum Fonden. The work at Tel Aviv Univesity was supported by ISF grants No. 1711/23 and 476/22 and by the Oren family Chair for experimental physics. Device fabrication was done with support from Technion's Micro-Nano Fabrication Unit (MNFU), with tremendous help from Amit Shacham, Dr. Orna Ternyak, and Dr. Tatiana Becker. The authors thank Dr. Michael Kalina for acquiring the optical image, and Dr. Anna Eyal for assistance with transport measurements. In addition, the authors sincerely acknowledge the contribution of Dr. Isao Inoue and thank him for fruitful discussions.

Supplementary Materials

**Charge gain via solid-state gating of an oxide Mott system**

*Lishai Shoham, Itai Silber, Gal Tuvia, Maria Baskin, Soo-Yoon Hwang, Si-Young Choi, Myung-Geun Han, Yimei Zhu, Eilam Yalon, Marcelo J. Rozenberg, Yoram Dagan, Felix Trier, and Lior Kornblum*

1. Film characterization

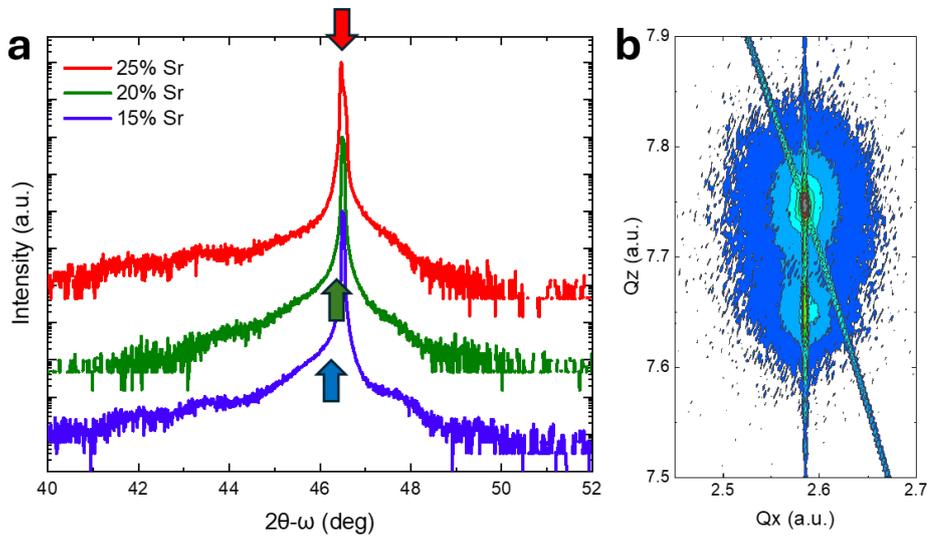

**Figure S1.** Structural analysis of the $La_{1-x}Sr_xVO_3$ (LSVO) films. (a) X-ray diffraction patterns around the cubic/pseuodocubic (002) Bragg reflection of ~4 nm LSVO films grown on $SrTiO_3$ (STO) substrates. The arrows indicate the location of the film's (002) peak as simulated using the Global Fit software. The scans were taken prior to the Hall bar patterning, from the same samples that were later used for the electrical measurements in the main text. (b) A reciprocal space map (RSM) scan around the (013) Bragg reflection of the STO substrate with a comparable but thicker 27 nm $La_{0.75}Sr_{0.25}VO_3$ film, showing coherent growth. All measurements were taken using a Rigaku SmartLab with a two-bounce incident monochromator.

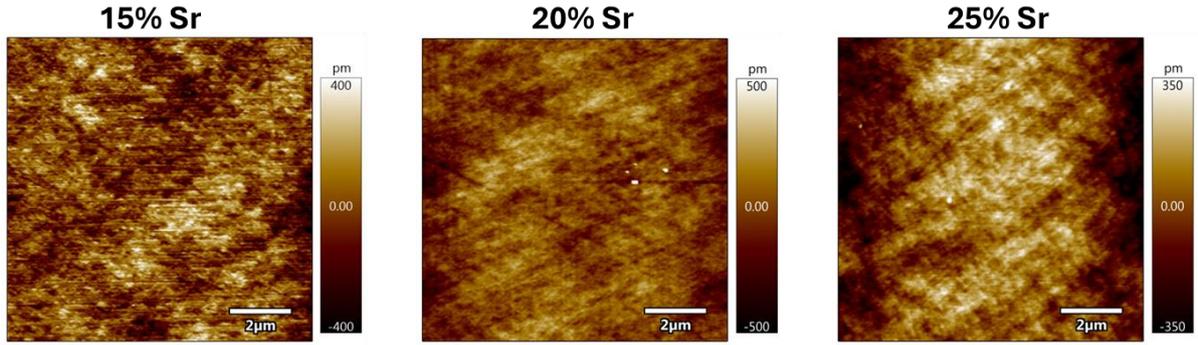

**Figure S2.** Surface morphology of the LSVO films prior to patterning illustrating a smooth surface (no substrate termination). The scans were acquired via an Asylum Research/Oxford Instruments Cypher ES Environmental atomic force microscope (AFM) operated in tapping mode.

## 2. Comparison of thick and thin films

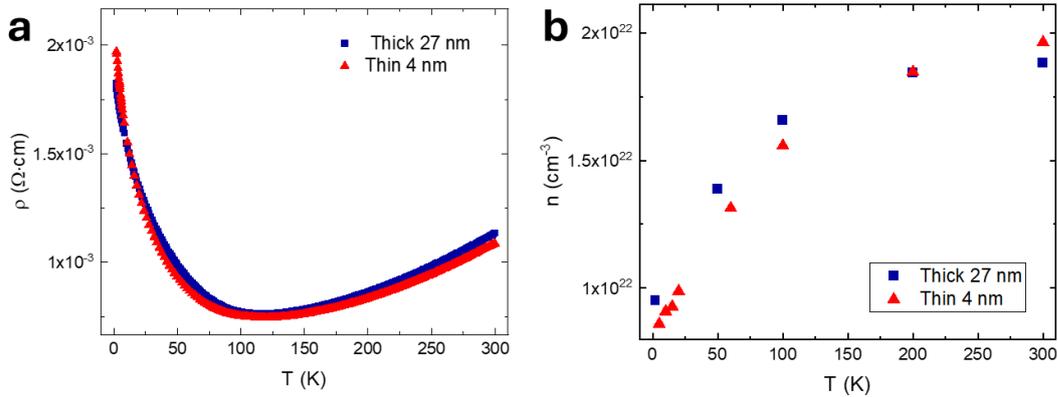

**Figure S3.** Magnetotransport comparison of $La_{0.75}Sr_{0.25}VO_3$ on STO substrate for Hall bar device (~4 nm) and a "thick" 27 nm film. (a) Temperature-dependent resistivity and (b) Temperature-dependent carrier density extracted from Hall measurements. For simplicity, we assume $n \approx n_H \approx (R_H e)^{-1}$ for a single-channel conduction (see discussion in the following section). The high resemblance of the results suggests a minor surface scattering in the Hall bar channel.

## 3. Mobility

Since the gate-induced charge variation of the LSVO Hall bar devices is negligible (see main text), we attribute the resistance response to the variation in mobility. Before attributing the mobility variation to changes in the electron correlations (manifest in the effective mass of the electrons, m*), we ruled out mobility degradation of the electrons under the gate electric field. Figure S4 presents the mobility of the LSVO devices at different temperatures, where the Hall response to the magnetic field was linear. The results illustrate that the mobility is relatively constant for the measured temperature range, and more importantly, decreases with the increase of electron density. Based on Matthiessen's rule, we expect that the mobility degradation under gate electric field would be more significant for the device with a higher mobility. However, comparing for example the resistance response between 20 and 25% Sr concentration at 10K (Figure S6), we can see that the resistance response of the 20% is about an order of magnitude higher than the response of the 25% device, thus ruling out significant mobility degradation in the LSVO devices.

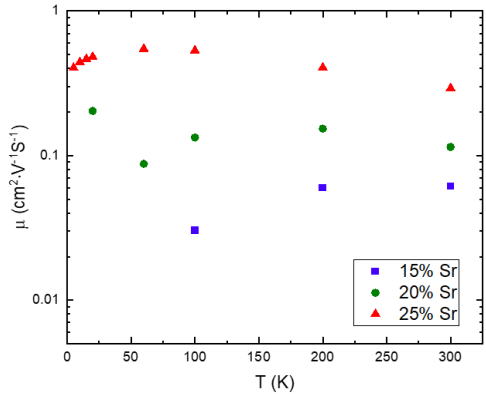

**Figure S4**. Temperature-dependent mobility for linear Hall response with variation of the magnetic field, assuming n $\approx (R_H e)^{-1}$.

## 4. Hall measurements

Hall measurements at various temperatures were conducted for all LSVO devices (Figure S5) in order to estimate the carrier density. For the 25% Sr concentration device, we report a linear Hall response with respect to the magnetic field for all temperatures measured, and only a small decrease in carrier concentration with cooling, which can be attributed to the LSVO paramagnetic metal phase.[1] Nevertheless, for the 20% Sr concentration device, we report a monotonically decline of the carrier density by an order of magnitude (or more) when cooling, which can be attributed to the LSVO antiferromagnetic metal phase.[1] In addition, for the 20% Sr concentration device, we observed a non-linear Hall response at low temperatures (< 20K). We attempted to fit a two-band model for the non-linear measurements, assuming a negligible 2-dimensional electron gas (2DEG) at the LSVO/STO interface. In this model, the electric conduction (σ) would be composed of a 3-dimensional (3D) contribution from the LSVO film and a 2D contribution from the 2DEG. The model used is based on the following:[2,3]

(1) $\sigma_{2D} = \begin{pmatrix} D_1 & -A_1 \\ A_1 & D_1 \end{pmatrix}$

(2) $D_1 = (n_1 \mu_1 q)/[1 + (\mu_1 B)^2]$

(3) $A_1 = D_1 \mu_1 B$

Where n and μ are the carrier density and mobility of the conduction channel, B is the external magnetic field, and q is the electron charge. Treating the 3D channel as three paths in parallel through the conduction band minimum (Γ, L and X) gives:

(4) $\sigma_i = \begin{pmatrix} D_i & -A_i \\ A_i & D_i \end{pmatrix}; i = \Gamma, L, X$

(5) $D_i = (n_i \mu_i q t)/[1 + (\mu_i B)^2]$

(6) $A_i = D_i \mu_i B$

And, by assuming an isotropic 3D conduction (Γ = L = X):

(7) $\sigma_{3D} = \sigma_\Gamma + \sigma_L + \sigma_X = 3\sigma$

(8) $\sigma_{tot} = \sigma_{3D} + \sigma_{2D}$

The results of the two-band model are summarized in Table 1. As can be seen, the fit to the model yielded unphysically low carrier density for the second conduction band (2DEG), suggesting that this model might be inadequate for our case. Another explanation for the non-linear Hall response may be the anomalous Hall effect, typically observed in ferromagnets but also possible in some antiferromagnets.[4] Recent theoretical work suggests that vanadates are candidates for anomalous Hall antiferromagnets due to their orbital ordering,[5,6] which links the two magnetic sublattices by rotation instead of translation, resulting in spin-split electronic bands. While this is not the focus of this work, it will be interesting to see whether future works could link the non-linear Hall effect we observe here with a magnetic origin. For the sake of clarity and focus, regions with more than single-channel conductivity were not included in the main text.

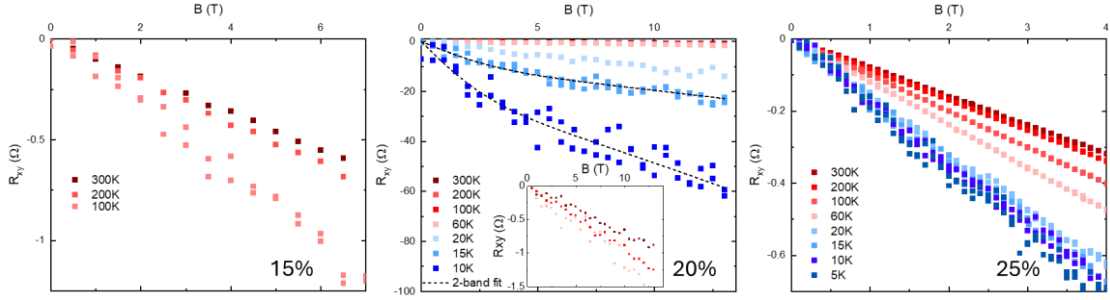

**Figure S5.** Hall measurements of the LSVO devices at different temperatures. To remove geometric errors, we measured the Hall voltage with both positive and negative fields and then subtracted one from the other: $R_{xy} = (R^{B+} - R^{B-})/2$.[7] When the Hall response is non-linear with respect to the magnetic field for the 20% Sr concentration device, a 2-band fit is presented based on equations 1-8 and the values in Table 1. At 50K and below, we were not able to acquire reliable data for the 15% Sr concentration device.

**Table S1**. Carrier density (n) and mobility (μ) for fitting the non-linear Hall behavior of the 20% Sr concentration device at low temperatures. $n_1$ and $\mu_1$ represent a possible 2D conduction at the LSVO/STO interface, and $n_2$ and $\mu_2$ represent the 3D conduction of the LSVO channel.

| T (K) | $n_1$ ($10^6$ cm$^{-3}$) | $n_2$ ($10^{20}$ cm$^{-3}$) | $\mu_1$ (cm$^2$V$^{-1}$S$^{-1}$) | $\mu_2$ (cm$^2$V$^{-1}$S$^{-1}$) |
|---|---|---|---|---|
| 15K | 7.1 | 3.7 | 1797 | 0.17 |
| 10K | 3.9 | 1.3 | 2072 | 0.27 |

## 5. Gate measurements at low temperatures

Gate measurements were conducted for all LSVO devices at low temperatures (Figure S6). As explained in the main text, we chose to focus on the 100 K measurements because at that temperature, the Hall response to the variation of the magnetic field is linear, and the LSVO is not expected to be close to any structural or magnetic phase transition.[1]

When reviewing the gate results in Figure S5, one must consider three different contributions to the behavior and magnitude of the resistance response to the gate voltage:

1. The dielectric constant of the gate insulator (STO) is considerably higher at low temperatures, which significantly increases the charge induced by the gate. In addition, at low temperatures (< 65K) the STO dielectric constant is affected by the applied electric field following the relation $\varepsilon_{STO} = \frac{1}{A+B \cdot E}$, where A and B are constants for a given temperature.[8]
2. For the 20 and 15% Sr concentration devices, there is a large decrease in carrier density with temperature.
3. The intrinsic correlated-electron behavior as discussed in the main text.

Since the three contributions are interrelated, it is difficult to distinguish their individual contributions.

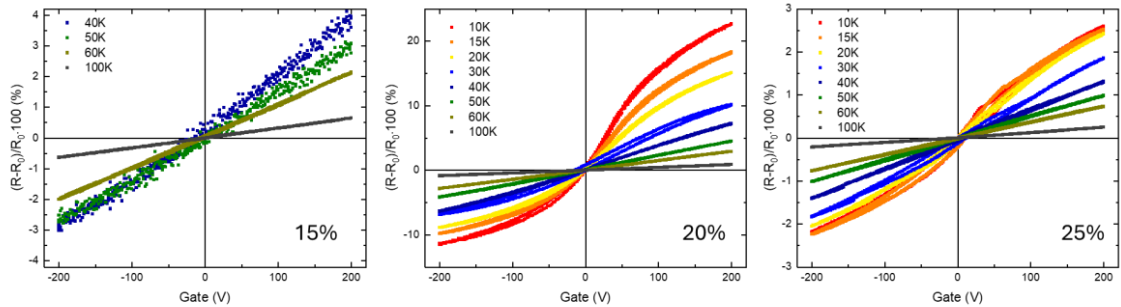

**Figure S6.** Resistance variation as a function of the gate voltage at different temperatures for different LSVO devices. $R_0$ is the device resistance under zero bias. The 15% Sr concentration device was too insulating to measure below 40 K.